# Fireball sheath instability

SUBHAM DUTTA and PRALAY KUMAR KARMAKAR[*]

Department of Physics, Tezpur University, Napaam-784 028, Tezpur, Assam, India.
[*]Corresponding author. E-mail: pkk@tezu.ernet.in

**Abstract.** The evolutionary existence of plasma fireballs is a generic phenomenon realisable in diversified physical plasma-dominated circumstances starting from the laboratory to the astrocosmic scales of space and time. A fair understanding of such fireballs and associated instabilities is indeed needed to enrich astroplasmic communities from various perspectives of applied value. Naturally occurring plasma fireball events include novae, meteors, stellar structures, etc. We propose a theoretical model formalism to analyse the plasma fireball sheath (PFS) instability with the application of a quasi-linear perturbative analysis on the laboratory spatiotemporal scales. This treatment reduces the steady-state system into a unique second-order ordinary differential equation (ODE) on the perturbed electrostatic potential with variable multiparametric coefficients. A numerical illustrative platform to integrate this ODE results in an atypical set of peakon-type potential-field structures. It is noticed that, both the potential and field associated with the peakonic patterns change significantly with the effective radial distance from the reference origin outwards. The variations are more pronounced at the centre (steep, stiff) than that in the off-centric regions (non-steep, non-stiff). A colormap obtained with the triangulation of the potential-field correlation with the radial distance further confirms the PFS stability behaviours in a qualitative corroboration with the previous predictions reported in the literature. The applicability of our analysis in both the laboratory and astrocosmic contexts is finally indicated.

## 1. Introduction

The interaction of plasma with embedded electrodes in the plasma chamber has been a topic of great interest over a few decades. It is seen that there exist extensive utilities of such plasma-wall effects in diversified cost-effective material processing and manufacturing techniques (Stenzel *et al*. 2021). It makes the plasma physics community to get more inclined towards understanding the basic physics behind the interaction processes involved in plasma-electrode (plasma-wall) interactions, instabilities, and various other related outcomes (Stenzel *et al*. 2008; Stenzel *et al*. 2021). Besides, such interaction processes result in the formation of non-neutral space charge layers, plasma sheaths, separating the bulk plasma from the rest.

It is seen that constitutive inter-species collision-induced avalanche-like expansion of plasma sheaths results in glowing structures (plasma fireballs). Such collisions lead to excitation of the neutral atoms. This excitation is responsible for the release of energy visible through the fireball spectral glow (Stenzel *et al.* 2008). The adjacent double layer (DL) outwards acts as a continuous source of charged particles developing the fireball. The fireball size relies on the rate of plasma production and of ion loss due to a modified electron-ion coupling (Stenzel *et al*. 2011). Also, the adjacent DL plays an active role in varying the size of the sheath, and thus, the fireball in the process. Clearly, fireballs are basically discharge phenomena, formed mostly in low-temperature partially ionized plasmas. The direct fireball forms in the sheath region around a solid electrode along with the formation of the DL outwards.

In addition to the above, another variety of fireballs, called inverted plasma fireballs, are formed with electrons converging inside a gridded hollow electrode. Similar to the fireballs, here too, the neutrals get excited and ionized due to collisions with fast moving electrons. The de-excitation process occurs with the energy released in the form of an inverted fireball structure (Gruenwald *et al.* 2014). In brief, we can state that a direct fireball forms around a continuous (regular) electrode (usually anode) surface; but, an inverted fireball forms inside a reticular (irregular) electrode (anode or cathode) volumetrically.

The practical realisation of the plasma fireball sheath (PFS) phenomena is quite extensive in the open areas of science, engineering, and technology (Stenzel *et al.* 2021). It evidently includes thermonuclear fusion plasmas, low-temperature plasmas, and plasma-assisted technology on various laboratory and astrocosmic spatiotemporal scales (Bandara & Khachan 2015; Harabasz *et al.* 2022). Besides, it helps in the formulation of high energy particle collision experiments generating quark-gluon plasma, especially the LHC, RHIC, and SPC. The fireball geometry and its expansion pattern help in describing both the yields and the spectra of various particles experimented upon at the Heavy Ion Synchrotron experiments, e.g., GSI SIS 18 in Germany and so forth (Harabasz *et al.* 2022). Using the fireball geometry and its expansion pattern Similarly, the study of inverted fireballs and the associated Buneman instabilities has significant importance in a good number of laboratory and astrophysical situations, such as stellar chromospheres (Fontenla *et al.* 2008), Supernova (SN2011fe) explosion (Dessart *et al.* 2014), cometary tails, bright meteoritic objects, ambient atmospheres, and so forth (https://www.imo.net/definitions-of-terms-in-meteor-astronomy-iau/). Application of the dynamic fireball model helps also in explaining astronomical expanding gamma-ray bursts effectively (Meng *et al.* 2022).

The fireballs appearing in astrocosmic environs are formed from the grape-sized meteors of adequate mass, which are, indeed, reduced gradually from the large massive asteroids. The characteristics of an astronomical fireball depend basically upon the threshold magnitude of luminosity required for normal observation and analysis. According to the recent reports, as published by the National Aeronautics and Space Administration (NASA), a meteor with an apparent brightness of magnitude at least -3 is technically said to be an astronomic fireball. Such fireballs are actually formed when a meteoroid (size ≈ a few cm) moving through the Earth's atmosphere at a hypersonic speed (11-72 km s$^{-1}$) compresses the surrounding air very rapidly. This compression results in heating the medium up to a very high degree (~1 kK). It radiates in the visible spectra in the mesospheric regions (altitude ≈ 60-90 km) (Bariselli *et al.* 2020). The observations of fireballs and environmental impacts show similar outcomes correlative with both the astronomical and laboratory plasmas. It includes the ionisation of the constitutive neutral components, sound wave generation, impulsive blast waves, diverse acousto-seismics, and so forth (Stenzel *et al.* 2008).

It is extensively evident that the PFS events are indeed spatiotemporally unstable on both the laboratory scales (Stenzel 1988; Stenzel *et al.* 2011) and astrocosmic scales (Meng *et al.* 2022). A theoretical model to explore the saturation mechanisms thereof is yet to be developed to the best of our knowledge. It forms the main motivational force to drive us towards this semi-analytic PFS analysis in a bifluidic fabric. A quasi-linear perturbative technique reduces the PFS system into a unique second-order ordinary differential equation (ODE) on the perturbed electrostatic potential with variable coefficients. A numerical integration (in Matlab) results in peakon-type structures, and so forth.

## 2. Physical model and formalism

We consider the formation of a PFS developed around a spherical electrode (anode) in a normal two-fluid plasma system (as in Figure 1). The proposed model consists mainly of electronic and ionic fluids with a negligible role of neutral component(s). It is obviously consistent and correlative with the plasmas well-realisable in laboratories. It may be noted herein that the experimental observations of plasma fireballs show a consistently spherical geometric shape in such circumstances (Stenzel *et al.* 2008). The consideration of the spherically symmetric geometry reduces the complex 3-D spherical problem into a simple 1-D radial problem without violating any kind of dynamical reality irrespective of spatiotemporal scales (Siemens & Rasmussen 1979). The full dynamics of the electronic and ionic fluids are governed by their respective electrodynamic governing equations evolving on their relevant laboratory spatiotemporal scales fulfilling the plasma existential conditions $((r,t) \gg (\lambda_D, \omega_p^{-1}))$. The model closure is obtained with the help of the coupling electrostatic Poisson equation describing the potential distribution developed due to local charge imbalance. The bulk macroscopic state of the entire plasma distribution forms a quasi-neutral hydrostatic homogenous equilibrium at least initially.

It may be noted herewith that the experimental observations of both plasma fireballs and related sheaths show diverse spatiotemporal fluctuations in terms of electrostatic potential, current, and emanated light (Stenzel *et al.* 2008). In our analysis, we consider only the asymptotic steady-state plasma fluctuations for the sake of simplifying the mathematical analysis against the initial state. Thus, it hereby enables us to transform our theoretic formalism to a steady-state model ansatz (time-stationary), but with no inherent loss of any kind of generality in the PFS evolution.

As already mentioned above, we are motivated to see a quasi-linear fireball sheath-plasma instability within the valid framework of a local perturbation analysis against the equilibrium. The corresponding basic governing equations describing the classical nonrelativistic dynamics of the PFS structure evolution with all the generic plasma notations (Gohain & Karmakar 2015) in a spherically symmetric coordination space $(r,t)$ are respectively enlisted as

$$\frac{\partial n_e}{\partial t} + \left(\frac{1}{r^2}\right)\frac{\partial}{\partial r}(r^2 n_e v_e) = 0, \tag{1}$$

$$m_e n_e \frac{\partial v_e}{\partial t} + m_e n_e v_e \frac{\partial v_e}{\partial r} = n_e e \frac{\partial \phi}{\partial r} - T_e \frac{\partial n_e}{\partial r}, \tag{2}$$

$$\frac{\partial n_i}{\partial t} + \left(\frac{1}{r^2}\right)\frac{\partial}{\partial r}(r^2 n_i v_i) = 0, \tag{3}$$

$$m_i n_i \frac{\partial v_i}{\partial t} + m_i n_i v_i \frac{\partial v_i}{\partial r} = n_i e \frac{\partial \phi}{\partial r} - T_i \frac{\partial n_i}{\partial r}. \tag{4}$$

Here, $n_{e(i)}$ denotes the electron (ion) population density, $v_{e(i)}$ denotes the electron (ion) flow velocity, $m_{e(i)}$ denotes the electron (ion) mass, and $T_{e(i)}$ is the electron (ion) temperature (in eV).

It is to be noted here that Equations (1)-(2) represent the equation of continuity and that of momentum for the electronic dynamics. Similarly, Equations (3)-(4) depict the same for the ions. The electron-ion closure electrostatic Poisson equation in the spherically symmetric geometry reads as

$$\frac{1}{r^2}\frac{\partial}{\partial r}\left(r^2 \frac{\partial \phi}{\partial r}\right) = 4\pi e(n_e - n_i). \tag{5}$$

Here, $\phi$ is the electrostatic potential developed due to local charge imbalance. The electron (ion) charge is $e = 1.9 \times 10^{-19}$ C is the magnitude of the. Clearly, it is revealed physically from Equation (5) that electric polarization effects (charge separation) always result in a commensurable electrostatic potential (or field) distribution in the coordination space irrespective of time scales. So, the left-hand side of Equation (5) is always independent of time, even if the charge distribution in the right-hand side in isolation is not. It well justifies the time-independent analytic nature of Equation (5).

We are interested in the steady-state evolution of the PFS fluctuations ($\partial/\partial t \sim 0$, but $\partial/\partial \xi \neq 0$) in the considered spherical geometry. We adopt here a standard normalization scheme well validated for laboratory plasmas (Karmakar & Gohain 2015). Accordingly, the normalized forms of Equations (1)-(4) are obtained in the time-stationary (steady-state) shape respectively as

$$M_e \frac{\partial N_e}{\partial \xi} + N_e \frac{\partial M_e}{\partial \xi} + \left(\frac{2}{\xi}\right) M_e N_e = 0, \tag{6}$$

$$N_e \frac{\partial \Phi}{\partial \xi} = N_e M_e \left(\frac{m_e}{m_i}\right)\frac{\partial M_e}{\partial \xi} + \frac{\partial N_e}{\partial \xi}, \tag{7}$$

$$M_i \frac{\partial N_i}{\partial \xi} + N_i \frac{\partial M_i}{\partial \xi} + \left(\frac{2}{\xi}\right) M_i N_i = 0, \tag{8}$$

$$N_i \frac{\partial \Phi}{\partial \xi} = N_i M_i \frac{\partial M_i}{\partial \xi} + \left(\frac{T_i}{T_e}\right)\frac{\partial N_i}{\partial \xi}. \tag{9}$$

The electrostatic Poisson equation (Equation (5)) in the normalized form is similarly cast as

$$\frac{\partial^2 \Phi}{\partial \xi^2} + \left(\frac{2}{\xi}\right)\frac{\partial \Phi}{\partial \xi} = N_e - N_i. \tag{10}$$

The normalized effective radial distance is given here as $\xi = r/\lambda_D$; where, $\lambda_D = \sqrt{T_e/4\pi n e^2}$ is the plasma Debye length. The normalized electron (ion) population density is $N_{e(i)} = n_{e(i)}/n_{e(o)} = n_{e(i)}/n_o$; where, $n_o$ is the equilibrium density. The electronic (ionic) fluid Mach number is $M_{e(i)} = v_{e(i)}/c_s$; where, $c_s = \sqrt{T_e/m_i}$ is the ion-sound phase speed. $\Phi = e\phi/T_e$ is the electrostatic potential normalized to the electron thermal potential, $T_e/e$.

The diverse relevant physical variables ($F_\alpha(\xi)$) in Equations (6)-(10) now undergo a quasi-linear local perturbation against their respective equilibrium values ($F_o$) on the $\epsilon$-order expansively as presented below

$$F(\xi) = F_o + \sum_{\alpha=1}^{\infty} \epsilon^\alpha F_\alpha(\xi). \tag{11}$$

Here, $\epsilon$ is an order parameter signifying the balanced strength of nonlinearity and dispersion (Mamun & Shukla 2002). An order-by-order analysis with Equation (11) put in Equations (6)-(9) up to the first-order yields

$$\frac{\partial M_{e1}}{\partial \xi} + \left(\frac{2}{\xi}\right) M_{e1} = 0, \tag{12}$$

$$\frac{\partial \Phi_1}{\partial \xi} = \frac{\partial N_{e1}}{\partial \xi}, \tag{13}$$

$$\frac{\partial M_{i1}}{\partial \xi} + \left(\frac{2}{\xi}\right) M_{i1} = 0, \tag{14}$$

$$\frac{\partial \Phi_1}{\partial \xi} = \left(\frac{T_i}{T_e}\right)\frac{\partial N_{i1}}{\partial \xi}. \tag{15}$$

After an indefinite $\xi$-integration on Equations (12)-(15) with the relevant boundary conditions (at $\xi = 0, \infty$) usually realizable in laboratory plasmas, we get the solutions of these equations respectively given as

$$M_{e1} = \frac{c_{eM}}{\xi^2}, \tag{16}$$

$$N_{e1} = \Phi_1 + c_{eN}, \tag{17}$$

$$M_{i1} = \frac{c_{iM}}{\xi^2}, \tag{18}$$

$$N_{i1} = \theta_{ei}\Phi_1 + c_{iN}. \tag{19}$$

Here, $c_{eM}$, $c_{eN}$, $c_{iM}$, and $c_{iN}$ are integration constants, which could be determined with the application of appropriate boundary conditions of the problem without loss of any generality. The electron-to-ion temperature ratio is $\theta_{ei} = T_e/T_i$. Replacement of $N_{e1}$ and $N_{i1}$ from respective Equation (17) and Equation (19) in the first-order quasi-linearly perturbed form of Equation (10) gives an ODE cast as

$$\frac{\partial^2 \Phi_1}{\partial \xi^2} + \left(\frac{2}{\xi}\right)\frac{\partial \Phi_1}{\partial \xi} = \Phi_1(1-\theta_{ei}) \\ -(c_{eN}-c_{iN}). \qquad (20)$$

which can be transformed as

$$\xi^2 \frac{\partial^2 \Phi_1}{\partial \xi^2} + 2\xi \frac{\partial \Phi_1}{\partial \xi} = \xi^2 \Phi_1(1-\theta_{ei}) \\ -\xi^2(c_{eN}-c_{iN}). \qquad (21)$$

The term, $(1-\theta_{ei})$, existent in Equations (20)-(21) physically denotes the deviation of the fireball-sheath region from the isothermal condition. The analytic solution of Equation (20), with the usual set of realistic boundary conditions (Chen 1984), manifests the evolution of the perturbed electrostatic potential in the PFS region. Matlab integration of Equation (20) yields

$$\Phi_1(\xi) = \frac{(1-(\sqrt{-c_1})\xi)c_2}{2(-c_1)^{\frac{3}{2}}\xi} - \frac{(1+(\sqrt{-c_1})\xi)c_2}{2(-c_1)^{\frac{3}{2}}\xi} \\ + \frac{\exp((\sqrt{-c_1})\xi)c_3}{\xi} - \frac{\exp(-(\sqrt{-c_1})\xi)c_4}{2(\sqrt{-c_1})\xi}. \qquad (22)$$

Here, $c_1 = \theta_{ei} - 1$, $c_2 = c_{eN} - c_{iN}$. $c_3$ and $c_4$ are the integration constants to be evaluated on the rounds of imposition of appropriate boundary conditions in conformity with bolstering experimental findings (Stenzel et al. 2021). The constant $c_2 = (c_{eN}-c_{iN})$, denotes the population density difference of the electrons and ions at a very large distance from the plasma sheath or the fireball. Where, the electrostatic potential due to the embedded electrode is not experienced. The constant $c_3$ denotes the electrostatic potential value at unit Debye length (See Equation (28)).

In nonthermal plasmas, $\theta_{ei} \geq 1$; so, one gets, $\sqrt{-c_1} = \sqrt{(1-\theta_{ei})} = i\sqrt{c_1}$. So, Equation (22) reads

$$\Phi_1(\xi) = \frac{(1-(i\sqrt{c_1})\xi)c_2}{2(-c_1)^{\frac{3}{2}}\xi} - \frac{(1+(i\sqrt{c_1})\xi)c_2}{2(-c_1)^{\frac{3}{2}}\xi} \\ + \frac{\exp((i\sqrt{c_1})\xi)c_3}{\xi} - \frac{\exp(-(i\sqrt{c_1})\xi)c_4}{2(i\sqrt{c_1})\xi}. \qquad (23)$$

A simplified form of Equation (23) can be given as

$$\Phi_1(\xi) = \frac{c_2}{c_1} + \frac{\exp((i\sqrt{c_1})\xi)c_3}{\xi} - \frac{\exp(-(i\sqrt{c_1})\xi)c_4}{2(i\sqrt{c_1})\xi}. \qquad (24)$$

It is noteworthy that the terms involving $1/\xi$ appears above because of spherical geometry; which would, otherwise, be absent on the grounds of planar geometry (Degasperis et al. 2002; Holm & Hone 2005). We are interested in the localized potential solutions around our reference point, $\xi \approx 0$. Application of the small amplitude approximation in Equation (24) gives

$$\Phi_1(\xi) = \frac{c_2}{c_1} + \frac{c_3}{\xi}\left[1 + i\sqrt{c_1}\,\xi + \frac{1}{2!}(i\sqrt{c_1})^2\xi^2 \right. \\ \left. + \frac{1}{3!}(i\sqrt{c_1})^3\xi^3 + \frac{1}{4!}(i\sqrt{c_1})^4\xi^4 + \cdots\right] \\ - \frac{c_4}{2(i\sqrt{c_1})\xi}\left[1 + (-i\sqrt{c_1})\xi + \frac{1}{2!}(-i\sqrt{c_1})^2\xi^2)\right. \\ \left. + \frac{1}{3!}(-i\sqrt{c_1})^3\xi^3 + \frac{1}{4!}(-i\sqrt{c_1})^4\xi^4 - \cdots\right]. \qquad (25)$$

Equation (25) can also be rearranged in terms of real and imaginary terms as given

$$\Phi_1(\xi) = \frac{c_2}{c_1} + \frac{c_3}{\xi}\left[(1 - \frac{1}{2!}c_1\xi^2 + \cdots) + i(\sqrt{c_1}\,\xi \right. \\ \left. - \frac{1}{3!}(c_1)^{\frac{3}{2}}\xi^3 + \cdots)\right] + \frac{c_4}{2\xi}\left[(\xi + \frac{1}{3!}c_1\xi^3 - \cdots) \right. \\ \left. - i((\sqrt{c_1})^{-1} - \frac{1}{2!}\sqrt{c_1}\xi^2 + \cdots)\right]. \qquad (26)$$

A comparison of the real and imaginary terms from both the left and the right-hand sides of Equation (26) yields the actual perturbed electrostatic potential as

$$\Phi_1(\xi) = \frac{c_2}{c_1} + \frac{c_3}{\xi}\left(1 - \frac{1}{2!}c_1\xi^2 + \frac{1}{4!}c_1^2\xi^4 - \cdots\right) \\ + \frac{c_4}{2\xi}\left(\xi - \frac{1}{3!}c_1\xi^3 + \frac{1}{5!}c_1^3\xi^5 - \cdots\right). \qquad (27)$$

It may be seen that Equation (27) is analogous to the common Fourier series expansion of any arbitrary signal (Arfken G. B. et al. 2012). In that Fourier perspective, the first free term, $c_2/c_1$ on the right-hand side would represent the zero-frequency line. The coefficient term, $c_3/\xi$, would denote the even harmonic strength. Similarly, the last coefficient term, $c_4/2\xi$, would designate the strength of odd harmonicity in the expansion series.

Using the physically sensible boundary conditions ($\Phi_1(\xi \to \pm\infty) \to 0$), Equation (27) yields

$$\Phi_1(\xi) = \frac{c_3}{\xi}. \qquad (28)$$

It may be noteworthy that, $c_3$ in Equation (28) could physically well signify the relative strength of the electrostatic potential also in the intervening plasma. Now, as we are interested only in the positive amplitude solutions of the locally perturbed electrostatic potential structures in full consistency with the literature (Stenzel et al. 2021), Equation (28) gets, therefore, amended to

$$\Phi_1(\xi) = \frac{c_3}{|\xi|}. \qquad (29)$$

The corresponding perturbed electric field, that is obtained from Equation (28) using the universal law of

conservative force field, can be presented as follows

$$E_1(\xi) = -\frac{\partial \Phi_1}{\partial \xi} = \frac{c_3}{\xi^2}. \quad (30)$$

The above analytic field expression, $E_1(\xi) = c_3/\xi^2$, can also be derived from the electrostatic potential, $\Phi_1(\xi) = c_3/\xi$, via a calculus limit theorem defined by the first principle as: $E_1 = -\underset{\xi_2 \to \xi_1}{Lt}[\Phi_1(\xi_2) - \Phi_1(\xi_1)]/(\xi_2 - \xi_1) = \underset{\xi_2 \to \xi_1}{Lt}[(c_3/\xi_1 - c_3/\xi_2)]/(\xi_2 - \xi_1) = c_3/\xi_2^2 = c_3/\xi^2$ for $\xi = \xi_2$. It could directly be used to examine the validity of the field-potential profile evolutionary mapping as elaborately discussed later.

A detailed numerical illustrative platform is developed to see the evolution of the potential perturbation in the form of spatial potential profiles ($\Phi_1 = \Phi_1(\xi)$, Figure 2), spatial field profiles ($E_1 = E_1(\xi)$, Figure 3), and conjoint colormaps with the help of a triangular coupling of Equation (29) and Equation (30) via $\xi$ ($E_1 = E_1(\xi, \Phi_1)$, Figure 4). The different colormaps link to the different $c_3$-values (as indicated).

## 3. Results and discussions

A fluid model is theoretically constructed herein in order to study the quasi-linear PFS instability dynamics on the laboratory spatiotemporal scales. The quasi-linearity arises because of a weak coupling between the nonlinear fluid convection and linear fluid dispersion (via $\epsilon$). Application of the quasi-linear perturbative analysis reduces the PFS system into a unique construct of a linear second-order ODE having a unique set of variable multiparametric coefficients (Equations (20)-(21)). The first-order perturbed electrostatic potential and the corresponding electric field are analytically evaluated and numerically portrayed in realistic plasma parametric windows. It is fairly supported by both the theoretical boundary conditions (Chen 1984) and the experimental PFS observations (Stenzel *et al.* 2021).

The peakon-type pulse potential structures are found to exist semi-analytically as the active signatures of the saturation process of the PFS instability (Figure 2). The associated narrow peakon fields are to be analyzed from a dispersive conservative force field perspective (Figure 3). A quicker reliability validation of Figure 3 may be instantly drawn by using the limiting graphical analysis in the light of Figure 2. It is seen that (Figure 2), $\Phi_1(\xi = -0.5) = 5$ and $\Phi_1(\xi = -0.6) = 4$; which gives $E_1 = [(\Phi_1(\xi = -0.5) - \Phi_1(\xi = -0.6)]/[-0.5 - (-0.6)] = 10$ (as in Figure 2). At the same time, one finds that, $E_1(\xi = -0.55) = 10$ graphically (Figure 3), which is the same as the above. It validates the reliability of our entire calculation scheme via Figure 3 as an outcome of Figure 2.

As in Figure 4, the colormaps generated in a color phase space (defined by a smooth triangulation of $\xi$, $\Phi_1$, and $E_1$) manifest the variation of the field strength with the variation of color density in accordance with Figures 2-3. Thus, the previous patterns (Figures 2-3) of the steady-state instability evolution are fairly confirmed by the color spectral profiles (Figure 4). In addition to the above, a spatially fast variation of the surface color around the reference point characterises a faster rate of the $E_1$-variation. In contrast, a slower color spectral variation asymptotically indicates a weaker change in $E_1$ from the reference point outwards. Thus, the steady saturation of the instability in the proposed plasma model configuration is graphically confirmed. It may be pertinent here to add further that the reliability of the spherically symmetric fireball model analysis is further strengthened in light of the qualitative matching of our obtained results (Figures 2-4) with the available experimental (Stenzel *et al.* 2021; Siemens & Rasmussen *et al.* 1979) and theoretical (Harabasz *et al.* 2022) findings reported extensively in the literature.

## 4. Conclusions

The steady PFS instability dynamics evolving around a spherical electrode realizable in the ambient plasmas is theoretically explored in a bifluidic model framework on the relevant laboratory scales of space and time. An applied quasi-linear perturbative analysis (relative to a well-defined hydrostatic homogeneous equilibrium) reduces the perturbed PFS system under test into a unique construct of a second-order linear ODE with variable coefficients. The formation of peakon-type potential and electric field structures around the electrode is numerically investigated. The peakonic features are confirmed in colorspectral phase space as well. It is shown that peakonic structures could result even from a linear ODE system against the traditional peakonic picture of fully nonlinear dynamical systems (Degasperis *et al.* 2002; Holm & Hone 2005). The representation of peakon structures herewith the exponential functions is quite in agreement with the earlier predictions reported in the literature (Degasperis *et al.* 2002; Holm & Hone 2005). As a consequence, it may be fairly conjectured that our analysis provides a theoretical platform to support the experimentally observed PFS potential structures from the bifluidic perspective. The atypical eigen-patterns are fairly consistent and correlative with experimental findings reported elsewhere (Stenzel *et al.* 2021). Thus, our analysis could be additionally applicable to understand the coupling stability scenarios of plasma sheath, fireball, and double layer on the usual laboratory scales.

This PFS instability analysis may indeed be efficacious in diversified plasma processing systems alongside the sheath-induced instability evolutionary phenomena of applied value in varied astrolabplasmic circumstances. In a broader sense, the plasma fireballs exert substantial pressure on the neutral and ionic components; thereby, inducing a macroscopic bulk gaseous flow in the test space-plasma medium taken under consideration, resulting in a number of plasma-jet phenomena. As a result, a fair understanding of the plasma fireballs and instabilities could enable us in developing a low-cost jet propulsion device for space-based technical explorations (Gruenwald *et al.* 2018). A comprehensive plasma concept of various associated instabilities is still in infancy stage as far as seen. It is believed that this analysis could be a promising element for illuminating this important direction having both laboratory and astrophysical plasma significances. Lastly, extended application of our dynamic fireball model in explaining astronomical expanding gamma-ray bursts effectively (Meng *et al.* 2022) could be another future scope of high explorative relevancy.

**Acknowledgements**

The active participation of the fellows of Astrophysical Plasma and Nonlinear Dynamics Laboratory, Tezpur University, is thankfully acknowledged. The valuable comments and suggestions received from the anonymous learned referees for scientific improvement are gratefully acknowledged. The financial support from the DST-SERB (Government of India) project (Grant-EMR/2017/003222) is duly recognized.

# Figures

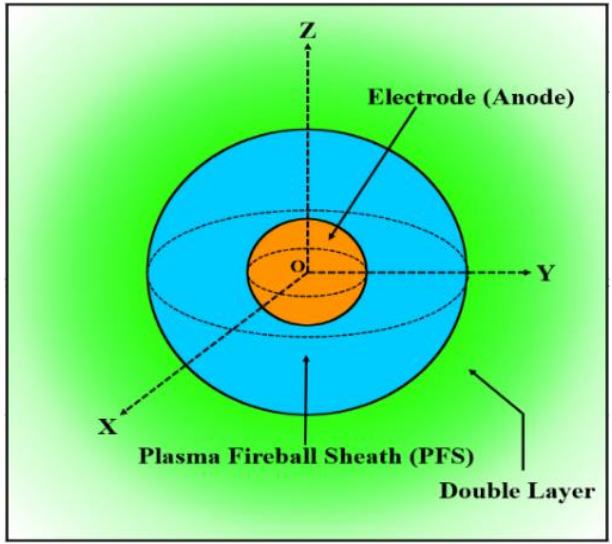

**Figure 1**. Schematic diagram showing the formation of plasma sheath, fireball, and double layer (DL) in a plasma chamber with an embedded spherical anode (central orange core). Its adjacent concentric spherical shell (sky blue region) displays the fireball sheath plasma (FSP). The circumvent outer region typifies the active zone for the DL formation (green region). The DL separates FSP from the bulk plasma. The origin of the coordinate system is assumed to lie at the anode surface.

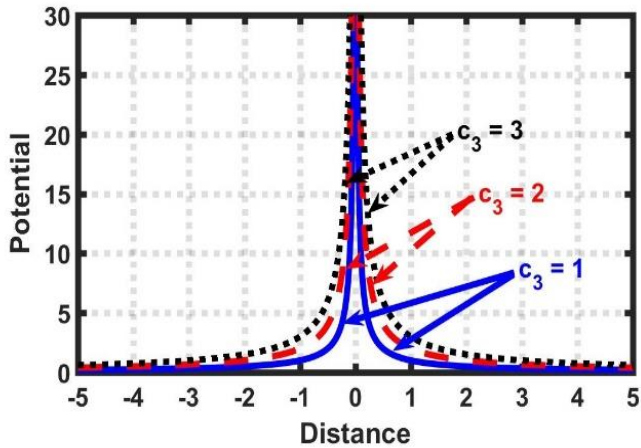

**Figure 2**. Profile of the perturbed normalized electrostatic potential ($\Phi_1 = f(\xi)$), governed by Equation (24)) with variation in the normalized radial distance relative to the electrode surface. The different lines link to the potential strength arising for $c_3 = 1$ (blue solid line), $c_3 = 2$ (red dashed line), and $c_3 = 3$ (black dotted line). The peakon-type structures peak at the origin of the spherical coordinate system.

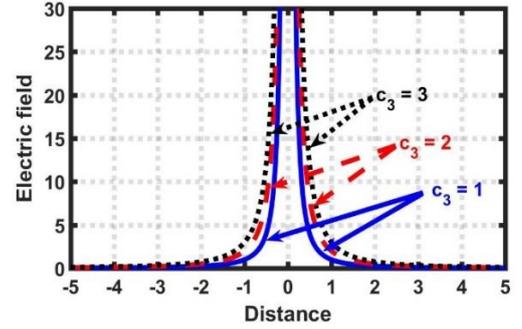

**Figure 3**. Same as Figure 2, but for the corresponding perturbed normalized electric field ($E_1 = -\partial \Phi_1/\partial \xi$, governed by Equation (25)).

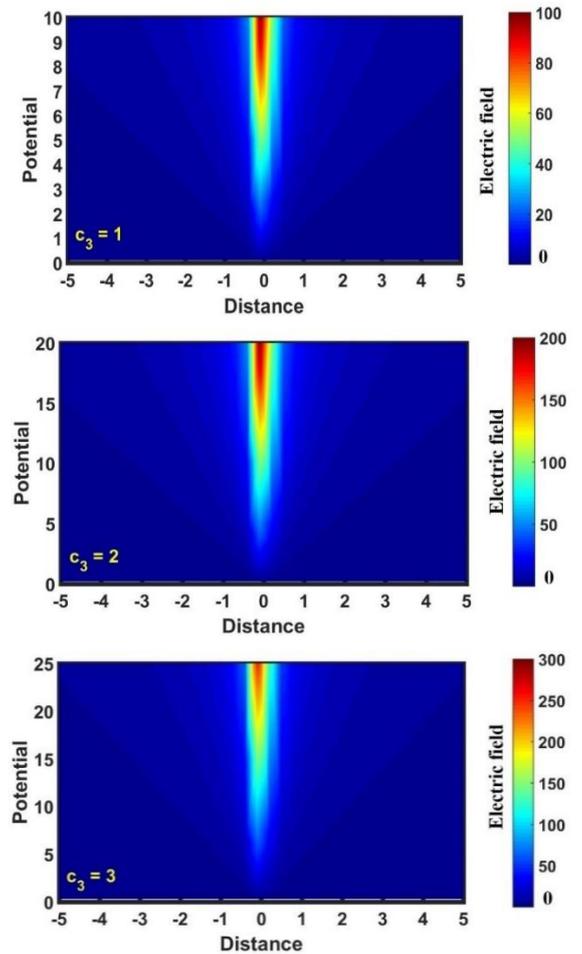

**Figure 4**. Colormap showing the evolution of the perturbed normalized electrostatic potential ($\Phi_1$) and the corresponding electric field ($E_1$) with variation in the normalized radial distance ($\xi$). The different maps correspond to $c_3 = 1, 2, 3$. A sharp color change at the center hints at a greater spatial rate of the ($\Phi_1, E_1$)-variation. The out-fading color replicates a meagre ($\Phi_1, E_1$)-change off-centrally.